\documentclass[preprint]{elsarticle}

\usepackage{amsmath,amsthm,amssymb,amsfonts}
\usepackage{dsfont}

\begin{document}

\begin{frontmatter}

\title{Finite-temperature second-order many-body perturbation theory revisited}

\author[desy,uhh]{Robin~Santra\corref{cor1}}
\ead{robin.santra@cfel.de}

\author[uhd]{Jochen~Schirmer}
\ead{jochen.schirmer@pci.uni-heidelberg.de}

\cortext[cor1]{Corresponding author}

\address[desy]{Center for Free-Electron Laser Science, DESY, Notkestr. 85, 22607 Hamburg, Germany}
\address[uhh]{Department of Physics, University of Hamburg, Jungiusstr. 9, 20355 Hamburg, Germany}
\address[uhd]{Theoretische Chemie, Physikalisch-Chemisches Institut, Universit\"{a}t Heidelberg,
Im Neuenheimer Feld 229, 69120 Heidelberg, Germany}

\begin{abstract}
We present an algebraic, nondiagrammatic derivation of finite-temperature se\-cond-order many-body perturbation 
theory [FT-MBPT(2)], using techniques and concepts accessible to theoretical chemical physicists. We give explicit 
expressions not just for the grand potential but particularly for the mean energy of an interacting many-electron 
system. The framework presented is suitable for computing the energy of a finite or infinite system in contact with 
a heat and particle bath at finite temperature and chemical potential. FT-MBPT(2) may be applied if the system, 
at zero temperature, may be described using standard (i.e., zero-temperature) second-order many-body perturbation 
theory [ZT-MBPT(2)] for the energy. We point out that in such a situation, FT-MBPT(2) reproduces, in the 
zero-temperature limit, the energy computed within ZT-MBPT(2). In other words, the difficulty that has been referred 
to as the Kohn--Luttinger conundrum, does not occur. We comment, in this context, on a ``renormalization'' scheme 
recently proposed by Hirata and He.
\end{abstract}

\end{frontmatter}

\section{\label{sec:1} Introduction}

Perturbation theory, particularly the variant due to Rayleigh and Schr\"{o}dinger \cite{Saku94,GoYa04}, is one of the 
most important approaches to finding approximate solutions to quantum-mechanical problems. In essence, one writes the 
Hamiltonian of interest, $\hat{H}$, as the sum of an unperturbed part, $\hat{H}_0$, and a perturbation, $\hat{H}_1$:
\begin{equation}
  \label{eq:0}
  \hat{H} = \hat{H}_0 + \hat{H}_1.
\end{equation}
Then, within time-independent perturbation theory, one constructs approximations to selected eigenstates and eigenenergies
of $\hat{H}$ using the eigenstates and eigenenergies of $\hat{H}_0$ (which must be known). The key assumption in the version
of time-independent perturbation theory that is suitable for nondegenerate states is that the zeroth-order reference state
is nondegenerate. When true degeneracies or quasi-degeneracies are present, then $\hat{H}$ must be prediagonalized within
the relevant (quasi-)degenerate subspace of $\hat{H}_0$ \cite{Kutz92,AnPa07,BrDu12}.

In practice, perturbation theory is most powerful when a low-order expansion suffices. The most widely used post-Hartree-Fock
method for the ground-state energy of an interacting many-electron system is second-order M{\o}ller--Plesset perturbation theory 
(MP2) \cite{MoPl34,KiKi98,Frie05,HeSc09}, which is second-order time-independent perturbation theory for nondegenerate states, employing 
an $\hat{H}_0$ that equals the ground-state Fock operator \cite{SzOs96} assuming a closed-shell system. MP2 works best when the 
ground-state Hartree-Fock HOMO--LUMO gap is, in some sense, not too small; MP2 diverges when the HOMO--LUMO gap 
vanishes \cite{GeBr57,GrMa10,ShGr13}.

MP2 is a special case of standard (i.e., zero-temperature) second-order many-body perturbation theory \cite{SzOs96,FeWa71,MaYo95,Wils07}, 
in the following referred to as ZT-MBPT(2). In this context, $\hat{H}$ is assumed to consist of one- and two-body operators \cite{FeWa71}. 
$\hat{H}$ is then partitioned such that 
\begin{equation}
  \label{eq:2a}
  \hat{H}_0 =  \sum_p \varepsilon_p \hat{c}^{\dag}_p \hat{c}_p
\end{equation}
is a one-body operator with known spin-orbital energies $\varepsilon_p$ and associated spin orbitals $\varphi_p$; 
$\hat{c}^{\dag}_p$ ($\hat{c}_p$) creates (annihilates) an electron in the one-electron state $\varphi_p$. Thus, the
perturbation
\begin{equation}
  \label{eq:2b}
  \hat{H}_1 = \sum_{p,q} v_{pq} \hat{c}^{\dag}_p \hat{c}_q 
  + \frac{1}{2} \sum_{p,q,r,s} v_{pqrs} \hat{c}^{\dag}_p \hat{c}^{\dag}_q \hat{c}_s \hat{c}_r
\end{equation}
generally consists of one- and two-body terms. In Eq.~(\ref{eq:2b}), $v_{pqrs}$ is an electron--electron Coulomb repulsion
integral. The one-electron integral $v_{pq}$ depends on the partitioning scheme selected; for instance, M{\o}ller--Plesset 
partitioning gives $v_{pq} = -\sum_{r} v_{pr[qr]} n_r$, where we have introduced the notation
\begin{equation}
  \label{eq:2c}
  v_{pr[qr]} = v_{prqr} - v_{prrq},
\end{equation}
and $n_r$ is the occupation number of spin orbital $\varphi_r$ in the Hartree-Fock ground state ($n_r = 0$ or $n_r = 1$).

The ZT-MBPT(2) result for the ground-state energy is \cite{SzOs96,Wils07}
\begin{eqnarray}
  \label{eq:2d}
  E_0 & = & E_0^{(0)} + E_0^{(1)} + E_0^{(2)}, \\ 
  \label{eq:2e}
  E_0^{(0)} & = & \sum_p \varepsilon_p n_p, \\
  \label{eq:2f}
  E_0^{(1)} & = & \sum_p v_{pp} n_p + \frac{1}{2} \sum_{p,q} v_{pq[pq]} n_p n_q, \\
  \label{eq:2g}
  E_0^{(2)} & = & -\sum_{p,q} \frac{n_q(1-n_p)}{\varepsilon_p - \varepsilon_q}
  \left|v_{pq} + \sum_{r} v_{pr[qr]} n_r\right|^2 \\
  & & - \frac{1}{4} \sum_{p,q,r,s}
  \frac{|v_{pq[rs]}|^2 n_r n_s (1-n_p) (1-n_q)}{\varepsilon_p + \varepsilon_q - \varepsilon_r - \varepsilon_s}. \nonumber
\end{eqnarray}
In the zero-temperature formalism, in contrast to the finite-temperature formalism that is in the focus of this paper,
the total particle number, $N = \sum_p n_p$, is a well-defined integer.

The traditional approach to extending many-body perturbation theory to finite temperature makes extensive use
of techniques adopted from quantum field theory \cite{FeWa71,MaYo95,Mats55,Thou57,BlDe58}. Particularly, there is an emphasis 
on diagrammatic techniques, using concepts that are not widely known in the theoretical chemical-physics community.
First steps towards introducing finite-temperature second-order many-body perturbation theory [FT-MBPT(2)]
to the chemical-physics literature were recently taken by Hirata and co-workers \cite{HiHe13,HiHe14}. Motivated
by the observation by Kohn and Luttinger \cite{KoLu60} that, in the zero-temperature limit, the mean energy
obtained within FT-MBPT(2) does not, in general, converge to the energy $E_0$ obtained within ZT-MBPT(2),
Hirata and co-workers proposed a ``renormalized'' version of FT-MBPT(2) \cite{HiHe13}.  

The present paper is an attempt to enhance the accessibility of finite-temperature many-body perturbation
theory through an elementary, nondiagrammatic derivation of FT-MBPT(2) equations for the mean energy and mean
particle number. These equations may be employed for describing finite or infinite electronic systems that are 
in contact with a heat and particle bath. This includes investigations of the electronic structure of
warm dense matter \cite{NeRe08,ViCi12,SoTh14,GlFl16}.

As we will show, if ZT-MBPT(2) is applicable, i.e., if there is a nonzero 
(ideally, large) HOMO--LUMO gap in the one-particle energy spectrum of $\hat{H}_0$, then, as the temperature
goes to zero, FT-MBPT(2) connects smoothly to ZT-MBPT(2). In other words, the Kohn--Luttinger conundrum, which
motivated the work of Hirata and co-workers \cite{HiHe13}, does not exist in situations in which the application of
second-order many-body perturbation theory is meaningful. Furthermore, we clarify in this paper the meaning
of what Hirata and co-workers call ``conventional'' FT-MBPT(2) \cite{HiHe13}. In contrast to what they suggested,
they did not, in fact, give an expression for the energy. Finally, we comment on their proposed ``renormalized''
FT-MBPT(2).

\section{\label{sec:2} Finite-temperature many-body perturbation theory}

Finite-temperature many-body perturbation theory (FT-MBPT) \cite{FeWa71,MaYo95,Mats55,Thou57,BlDe58} is based 
on the grand-canonical ensemble \cite{Huan87}. The fundamental quantity describing the state of a system in the 
grand-canonical ensemble, such that the parameters of the theory are the temperature $T$ (or $\beta = 1/T$ in suitable units), 
the volume $V$, and the chemical potential $\mu$, is the density operator 
\begin{equation}
  \label{eq:1}
  \hat{\rho} = \frac{e^{-\beta(\hat{H}- \mu \hat{N})}}{Z_G}.
\end{equation}
Here,
\begin{equation}
  \label{eq:3}
  \hat{N} = \sum_p \hat{c}^{\dag}_p \hat{c}_p
\end{equation}
is the total particle number operator,
and 
\begin{equation}
  \label{eq:4}
  Z_G = \mathrm{Tr}\left\{e^{-\beta(\hat{H}- \mu \hat{N})}\right\}
\end{equation}
is the grand partition function.

For the noninteracting reference system, the grand-canonical density operator is given by
\begin{equation}
  \label{eq:5}
  \hat{\rho}_0 = \frac{e^{-\beta(\hat{H}_0- \mu \hat{N})}}{Z_G^{(0)}},
\end{equation}
where
\begin{equation}
  \label{eq:6}
  Z_G^{(0)} = \mathrm{Tr}\left\{e^{-\beta(\hat{H}_0- \mu \hat{N})}\right\}.
\end{equation}
The Fermi-Dirac factor, 
\begin{equation}
  \label{eq:7}
  \bar{n}_p = \frac{1}{e^{\beta(\varepsilon_p - \mu)} + 1},
\end{equation}
emerges, for the noninteracting reference system, as the ensemble-averaged expectation value of 
the spin-orbital particle number operator 
\begin{equation}
  \label{eq:8}
  \hat{n}_p = \hat{c}^{\dag}_p \hat{c}_p,
\end{equation}
i.e.,
\begin{equation}
  \label{eq:9}
  \bar{n}_p = \langle \hat{n}_p \rangle_0 = \mathrm{Tr}\left\{\hat{\rho}_0 \hat{n}_p\right\}.
\end{equation}
As is well known, this is the sole meaning of the Fermi-Dirac factor in Eq.~(\ref{eq:7}). It is not a fundamental quantity of 
quantum statistical mechanics; it is derived from Eq.~(\ref{eq:9}) using Eqs.~(\ref{eq:5}) and (\ref{eq:6}).

FT-MBPT is not a perturbation theory directly for the (mean) energy of a given system,
but for its grand partition function, Eq.~(\ref{eq:4}). To this end, note that the operator
\begin{equation}
  \label{eq:9a}
  \hat{U}(\beta) = e^{-\beta(\hat{H} - \mu \hat{N})}
\end{equation}
appearing in Eq.~(\ref{eq:4}) has the structure of a time evolution operator with time argument $-i\beta$ (``imaginary time'') 
and Hamiltonian $\hat{H} - \mu \hat{N}$. One can, thus, define a corresponding operator in the interaction picture,
\begin{equation}
  \label{eq:9b}
  \hat{U}_{\mathrm{I}}(\beta) = e^{\beta(\hat{H}_0 - \mu \hat{N})} \hat{U}(\beta).
\end{equation}
This satisfies the ``equation of motion'' (known as Bloch equation)
\begin{equation}
  \label{eq:9c}
  \frac{\partial}{\partial\beta} \hat{U}_{\mathrm{I}}(\beta) = - \hat{H}_1(\beta) \hat{U}_{\mathrm{I}}(\beta),
\end{equation}
where
\begin{equation}
  \label{eq:9d}
  \hat{H}_1(\beta) = e^{\beta(\hat{H}_0- \mu \hat{N})} \hat{H}_1 e^{-\beta(\hat{H}_0- \mu \hat{N})}.
\end{equation}

At $\beta = 0$, i.e., at infinite temperature, $\hat{U}=\mathds{1}$ (the identity operator) and, therefore,
$\hat{U}_{\mathrm{I}}=\mathds{1}$. Using this point of reference, Eq.~(\ref{eq:9c}) may be integrated and the
resulting integral equation may be solved iteratively. Hence, through second order we have
\begin{eqnarray}
  \label{eq:9e}
  \hat{U}_{\mathrm{I}}(\beta) & = & \mathds{1} - \int_0^{\beta} du \, \hat{H}_1(u) \\
  & & + \int_0^{\beta} du \, \hat{H}_1(u) \int_0^{u} du' \, \hat{H}_1(u'). \nonumber
\end{eqnarray}
In analogy to Eq.~(\ref{eq:9d}),
\begin{equation}
  \label{eq:10b}
  \hat{H}_1(u) = e^{u(\hat{H}_0- \mu \hat{N})} \hat{H}_1 e^{-u(\hat{H}_0- \mu \hat{N})}.
\end{equation}
By combining Eq.~(\ref{eq:9e}) with Eqs.~(\ref{eq:4}), (\ref{eq:5}), (\ref{eq:6}), (\ref{eq:9a}), and (\ref{eq:9b}), it follows that
\begin{eqnarray}
  \label{eq:10}
  \frac{Z_G}{Z_G^{(0)}} & = & 1 - \int_0^{\beta} du \, \mathrm{Tr}\left\{\hat{\rho}_0 \hat{H}_1(u)\right\} \\
  & & + \int_0^{\beta} du \int_0^{u} du' \, \mathrm{Tr}\left\{\hat{\rho}_0 \hat{H}_1(u) \hat{H}_1(u')\right\} + \ldots. \nonumber
\end{eqnarray}

Quantities such as the grand potential and the mean energy are more closely connected to $\ln{Z_G}$ than to $Z_G$ itself. 
(The grand potential is also known as the thermodynamic potential \cite{FeWa71}.) Using
\begin{equation}
  \label{eq:11}
  \ln{(1+x)} = x - \frac{x^2}{2} + \ldots,
\end{equation}
we obtain from Eq.~(\ref{eq:10}) a perturbation expansion for $\ln{Z_G}$, valid through second order:
\begin{eqnarray}
  \label{eq:12}
  \ln{Z_G} & = & \ln{Z_G^{(0)}} - \int_0^{\beta} du \, \mathrm{Tr}\left\{\hat{\rho}_0 \hat{H}_1(u)\right\} \\
  & & + \int_0^{\beta} du \int_0^{u} du' \, \mathrm{Tr}\left\{\hat{\rho}_0 \hat{H}_1(u) \hat{H}_1(u')\right\} \nonumber \\
  & & - \frac{1}{2} \int_0^{\beta} du \int_0^{\beta} du' \, \mathrm{Tr}\left\{\hat{\rho}_0 \hat{H}_1(u)\right\}
  \mathrm{Tr}\left\{\hat{\rho}_0 \hat{H}_1(u')\right\}. \nonumber
\end{eqnarray}

\section{\label{sec:3} Nondiagrammatic evaluation of $\ln{Z_G}$}

The zeroth-order term on the right-hand side of Eq.~(\ref{eq:12}), i.e., the natural logarithm of the grand partition 
function of a system of effectively noninteracting fermions described by the Hamiltonian $\hat{H}_0$ [Eq.~(\ref{eq:2a})],
may be found in virtually any textbook on statistical mechanics (see, for example, Ref.~\cite{Huan87}):
\begin{equation}
  \label{eq:20}
  \ln{Z_G^{(0)}} = \sum_p \ln{\left(1 + e^{-\beta(\varepsilon_p - \mu)}\right)}.
\end{equation}

An eigenstate $|\{n\}\rangle$ of $\hat{H}_0$ and $\hat{N}$ [Eq.~(\ref{eq:3})] is a Fock state characterized by a specific 
spin-orbital occupation pattern ($\{n\} = n_1, n_2, \ldots$ is a collective index, where each occupation number $n_p$ 
equals either $0$ or $1$). Using $\hat{c}^{\dag}_p \hat{c}_p|\{n\}\rangle = n_p|\{n\}\rangle$, the eigenvalues of $\hat{H}_0$ 
and $\hat{N}$ associated with the Fock state $|\{n\}\rangle$ read, respectively,
\begin{eqnarray}
  \label{eq:12a}
  E_{\{n\}}^{(0)} & = & \sum_p \varepsilon_p n_p, \\
  \label{eq:12b}
  N_{\{n\}} & = & \sum_p n_p. 
\end{eqnarray}

Thus, with the aid of the Slater-Condon rules \cite{SzOs96}, applied to $\hat{H}_1$ as given in Eq.~(\ref{eq:2b}),
we may evaluate the first-order term in Eq.~(\ref{eq:12}) as follows:
\begin{eqnarray}
  \label{eq:12c}
  - \int_0^{\beta} du \, \mathrm{Tr}\left\{\hat{\rho}_0 \hat{H}_1(u)\right\} & = & - \int_0^{\beta} du \, \mathrm{Tr}\left\{\hat{\rho}_0 \hat{H}_1\right\} \\
  & = & - \beta \, \mathrm{Tr}\left\{\hat{\rho}_0 \hat{H}_1\right\} \nonumber \\
  & = & - \beta \sum_{\{n\}} \frac{e^{-\beta\left(E_{\{n\}}^{(0)} - \mu N_{\{n\}}\right)}}{Z_G^{(0)}} \langle\{n\}|\hat{H}_1|\{n\}\rangle \nonumber \\
  & = & - \beta \sum_{\{n\}} \frac{e^{-\beta\left(E_{\{n\}}^{(0)} - \mu N_{\{n\}}\right)}}{Z_G^{(0)}} \nonumber \\
  & & \times \left\{\sum_p v_{pp} n_p + \frac{1}{2} \sum_{p,q} v_{pq[pq]} n_p n_q\right\} \nonumber \\
  & = & - \beta \sum_{\{n\}} \frac{e^{-\beta\left(E_{\{n\}}^{(0)} - \mu N_{\{n\}}\right)}}{Z_G^{(0)}} \nonumber \\
  & & \times \left\{\sum_p v_{pp} \langle\{n\}|\hat{n}_p|\{n\}\rangle 
    + \frac{1}{2} \sum_{p,q} v_{pq[pq]} \langle\{n\}|\hat{n}_p\hat{n}_q|\{n\}\rangle\right\} \nonumber \\
  & = & - \beta \left\{\sum_p v_{pp} \langle\hat{n}_p\rangle_0 + \frac{1}{2} \sum_{p,q} v_{pq[pq]} \langle\hat{n}_p\hat{n}_q\rangle_0\right\}. \nonumber
\end{eqnarray}
In the first line of Eq.~(\ref{eq:12c}) we made use of the cyclic property of the trace.
The expectation value $\langle\hat{n}_p\rangle_0$ in the last line of Eq.~(\ref{eq:12c}) is simply a Fermi-Dirac factor $\bar{n}_p$
[see Eqs.~(\ref{eq:7}) and (\ref{eq:9})]. More generally, as in the second term in the curly braces in the last line of Eq.~(\ref{eq:12c}),
one must use 
\begin{equation}
  \label{eq:12e}
  \langle\hat{n}_p^{m_p}\hat{n}_q^{m_q}\ldots\rangle_0 = \bar{n}_p \bar{n}_q \ldots,
\end{equation}
where the $p$, $q$, ... are all different from each other, and the $m_p$, $m_q$, ... are positive integers. 
[Proving Eq.~(\ref{eq:12e}) is a simple exercise.] Hence, the first-order term in Eq.~(\ref{eq:12}) is given by
\begin{equation}
  \label{eq:12f}
  - \int_0^{\beta} du \, \mathrm{Tr}\left\{\hat{\rho}_0 \hat{H}_1(u)\right\} = 
  - \beta \left\{\sum_p v_{pp} \bar{n}_p + \frac{1}{2} \sum_{p,q} v_{pq[pq]} \bar{n}_p \bar{n}_q \right\}.
\end{equation}

In order to evaluate the first second-order term in Eq.~(\ref{eq:12}), we proceed as follows:
\begin{multline}
  \label{eq:12g}
    \int_0^{\beta} du \int_0^{u} du' \, \mathrm{Tr}\left\{\hat{\rho}_0 \hat{H}_1(u) \hat{H}_1(u')\right\} 
    =  \int_0^{\beta} du \int_0^{u} du' \, \sum_{\{n\},\{n'\}} \frac{e^{-\beta\left(E_{\{n\}}^{(0)} - \mu N_{\{n\}}\right)}}{Z_G^{(0)}} \\
    \times e^{u\left(E_{\{n\}}^{(0)} - E_{\{n'\}}^{(0)}\right)}\langle\{n\}|\hat{H}_1|\{n'\}\rangle 
    e^{u'\left(E_{\{n'\}}^{(0)} - E_{\{n\}}^{(0)}\right)}\langle\{n'\}|\hat{H}_1|\{n\}\rangle \\
    = \sum_{\{n\},\{n'\}} \frac{e^{-\beta\left(E_{\{n\}}^{(0)} - \mu N_{\{n\}}\right)}}{Z_G^{(0)}} 
     \frac{\left|\langle\{n'\}|\hat{H}_1|\{n\}\rangle\right|^2}{E_{\{n'\}}^{(0)} - E_{\{n\}}^{(0)}} 
     \left\{\beta - \frac{e^{\beta\left(E_{\{n\}}^{(0)} - E_{\{n'\}}^{(0)}\right)}-1}{E_{\{n\}}^{(0)} - E_{\{n'\}}^{(0)}}\right\}.
\end{multline}
Despite its appearance, energy degeneracies cause no difficulties in this expression (for finite $\beta$):
\begin{equation}
  \label{eq:12h}
  \lim_{E_{\{n'\}}^{(0)} \rightarrow E_{\{n\}}^{(0)}} \frac{\left|\langle\{n'\}|\hat{H}_1|\{n\}\rangle\right|^2}{E_{\{n'\}}^{(0)} - E_{\{n\}}^{(0)}} 
     \left\{\beta - \frac{e^{\beta\left(E_{\{n\}}^{(0)} - E_{\{n'\}}^{(0)}\right)}-1}{E_{\{n\}}^{(0)} - E_{\{n'\}}^{(0)}}\right\}
     = \frac{1}{2}\beta^2 \left|\langle\{n'\}|\hat{H}_1|\{n\}\rangle\right|^2.
\end{equation}

Since $\hat{H}_1$ consists only of one- and two-body operators, for any given $|\{n\}\rangle$ in Eq.~(\ref{eq:12g}), only those Fock states $|\{n'\}\rangle$
can make a nonzero contribution that fall into one of the following three categories: (i) $|\{n'\}\rangle = |\{n\}\rangle$;
(ii) $|\{n'\}\rangle = \hat{c}^{\dag}_p \hat{c}_q|\{n\}\rangle$ with $n_p = 0$ and $n_q = 1$ (in the notation adopted here, the occupation
numbers refer to the Fock state $|\{n\}\rangle$); (iii) $|\{n'\}\rangle = \hat{c}^{\dag}_p \hat{c}^{\dag}_q \hat{c}_s \hat{c}_r |\{n\}\rangle$ with 
$n_p = 0$, $n_q = 0$, $n_r = 1$, and $n_s = 1$ (and, of course, $p\ne q$ and $r \ne s$). In each of these cases, we may employ the corresponding 
Slater-Condon rules to reduce the matrix elements $\langle\{n'\}|\hat{H}_1|\{n\}\rangle$ to one- and two-electron integrals multiplied by suitable 
occupation numbers. For example, if $|\{n'\}\rangle$ falls into category (ii), then 
$\langle\{n'\}|\hat{H}_1|\{n\}\rangle = \left(v_{pq} + \sum_{r} v_{pr[qr]} n_r\right)n_q(1-n_p)$.
After steps analogous to those shown in Eq.~(\ref{eq:12c}), we obtain from Eq.~(\ref{eq:12g}) expressions that depend on expectation values of the form
$\langle\hat{n}_p^{m_p}\hat{n}_q^{m_q}\ldots\rangle_0$, which may be reduced to products of Fermi-Dirac factors using Eq.~(\ref{eq:12e}).

From the perspective of ZT-MBPT(2), it is tempting to expect that the contributions from category (i) are cancelled by the 
second second-order term in Eq.~(\ref{eq:12}). This, however, is not completely the case:
\begin{multline}
  \label{eq:12i}
  \int_0^{\beta} du \int_0^{u} du' \, \sum_{\{n\}} \frac{e^{-\beta\left(E_{\{n\}}^{(0)} - \mu N_{\{n\}}\right)}}{Z_G^{(0)}} 
  \left|\langle\{n\}|\hat{H}_1|\{n\}\rangle\right|^2 \\
  -\frac{1}{2} \int_0^{\beta} du \int_0^{\beta} du' \, \mathrm{Tr}\left\{\hat{\rho}_0 \hat{H}_1(u)\right\}
  \mathrm{Tr}\left\{\hat{\rho}_0 \hat{H}_1(u')\right\} \\
  = \frac{1}{2} \beta^2 \left\{\sum_p\left|v_{pp} + \sum_{q} v_{pq[pq]} \bar{n}_q\right|^2 \bar{n}_p(1-\bar{n}_p)
    + \frac{1}{2} \sum_{p,q} \left|v_{pq[pq]}\right|^2\bar{n}_p(1-\bar{n}_p) \bar{n}_q(1-\bar{n}_q)\right\}. 
\end{multline}
The second-order contributions from categories (ii) and (iii) read: 
\begin{multline}
  \label{eq:12j}
  \sum_{\{n\}} \sum_{\{n'\}\ne\{n\}} \frac{e^{-\beta\left(E_{\{n\}}^{(0)} - \mu N_{\{n\}}\right)}}{Z_G^{(0)}}
     \frac{\left|\langle\{n'\}|\hat{H}_1|\{n\}\rangle\right|^2}{E_{\{n'\}}^{(0)} - E_{\{n\}}^{(0)}}
     \left\{\beta - \frac{e^{\beta\left(E_{\{n\}}^{(0)} - E_{\{n'\}}^{(0)}\right)}-1}{E_{\{n\}}^{(0)} - E_{\{n'\}}^{(0)}}\right\} \\
= \sum_{p,q} \frac{\bar{n}_q(1-\bar{n}_p)}{\varepsilon_p - \varepsilon_q}
  \left|v_{pq} + \sum_{r} v_{pr[qr]} \bar{n}_r\right|^2
  \left\{\beta - \frac{1-e^{-\beta(\varepsilon_p - \varepsilon_q)}}{\varepsilon_p - \varepsilon_q}\right\} \\
  - \frac{1}{2} \beta^2 \left\{\sum_p\left|v_{pp} + \sum_{q} v_{pq[pq]} \bar{n}_q\right|^2 \bar{n}_p(1-\bar{n}_p)
  + \frac{1}{2} \sum_{p,q} \left|v_{pq[pq]}\right|^2\bar{n}_p(1-\bar{n}_p) \bar{n}_q(1-\bar{n}_q)\right\} \\
  + \frac{1}{4} \sum_{p,q,r,s} \frac{|v_{pq[rs]}|^2 \bar{n}_r \bar{n}_s (1-\bar{n}_p) (1-\bar{n}_q)}{\varepsilon_p +
    \varepsilon_q - \varepsilon_r - \varepsilon_s}
  \left\{\beta -
    \frac{1-e^{-\beta(\varepsilon_p + \varepsilon_q - \varepsilon_r - \varepsilon_s)}}{\varepsilon_p + \varepsilon_q -
      \varepsilon_r - \varepsilon_s}\right\}.
\end{multline}
Hence, the second term on the right-hand side of Eq.~(\ref{eq:12j}) finally cancels the residual contribution from Eq.~(\ref{eq:12i}).

Overall, using the elementary, nondiagrammatic approach described, we find for the natural logarithm of the grand partition function through
second order:
\begin{eqnarray}
  \label{eq:13}
  \ln{Z_G} & = & \sum_p \ln{\left(1 + e^{-\beta(\varepsilon_p - \mu)}\right)} \\
  & & -\beta \left(\sum_p v_{pp} \bar{n}_p + \frac{1}{2} \sum_{p,q} v_{pq[pq]} \bar{n}_p \bar{n}_q\right) \nonumber \\
  & &  + \sum_{p,q} \frac{\bar{n}_q(1-\bar{n}_p)}{\varepsilon_p - \varepsilon_q} 
  \left|v_{pq} + \sum_{r} v_{pr[qr]} \bar{n}_r\right|^2 
  \left\{\beta - \frac{1-e^{-\beta(\varepsilon_p - \varepsilon_q)}}{\varepsilon_p - \varepsilon_q}\right\} \nonumber \\
  & & + \frac{1}{4} \sum_{p,q,r,s} \frac{|v_{pq[rs]}|^2 \bar{n}_r \bar{n}_s (1-\bar{n}_p) (1-\bar{n}_q)}{\varepsilon_p + 
    \varepsilon_q - \varepsilon_r - \varepsilon_s}
  \left\{\beta - 
    \frac{1-e^{-\beta(\varepsilon_p + \varepsilon_q - \varepsilon_r - \varepsilon_s)}}{\varepsilon_p + \varepsilon_q - 
      \varepsilon_r - \varepsilon_s}\right\}. \nonumber
\end{eqnarray}

\section{\label{sec:4} The Kohn--Luttinger conundrum}

In order to calculate the mean energy of the interacting system,
\begin{equation}
  \label{eq:14}
  E = \langle \hat{H} \rangle = \mathrm{Tr}\left\{\hat{\rho} \hat{H}\right\}, 
\end{equation}
one option is to recognize that the grand potential 
\begin{equation}
  \label{eq:15}
  \Omega = E - TS - \mu \langle \hat{N} \rangle
\end{equation}
is connected to the grand partition function via
\begin{equation}
  \label{eq:16}
  Z_G = e^{-\beta\Omega}, 
\end{equation}
such that
\begin{equation}
  \label{eq:17}
  E = -\frac{1}{\beta} \ln{Z_G} + TS + \mu \langle \hat{N} \rangle.
\end{equation}
This is the approach employed by Kohn and Luttinger \cite{KoLu60}, who focused exclusively on the limit $T \rightarrow 0$ 
and exploited that, in this limit, $TS \rightarrow 0$
(so they did not have to determine an expression for the entropy $S$ of the system). 

The essence of what Hirata and He \cite{HiHe13} have referred to as the Kohn--Luttinger conundrum is that, generally, 
$E$ using Eqs.~(\ref{eq:13}) and (\ref{eq:17}) does not converge, in the 
zero-temperature limit, to the energy one would obtain in ZT-MBPT(2). This is easy to understand by inspecting the third line of 
Eq.~(\ref{eq:13}). The sum over $p$ and $q$ includes terms where $p=q$. By letting $\varepsilon_p - \varepsilon_q \rightarrow 0$ at finite $\beta$, 
the contribution of those $p=q$ terms to the energy $E$ is found to be 
\[-\frac{\beta}{2}\sum_{p} \bar{n}_p(1-\bar{n}_p) \left|v_{pp} + \sum_{r} v_{pr[pr]} \bar{n}_r\right|^2.\]
This corresponds to $\Omega_{2A}$ in the paper by Kohn and Luttinger \cite{KoLu60} (except that we include in $\hat{H}_1$ a one-body operator, giving rise to $v_{pp}$,
and we don't assume that the direct Coulomb matrix element $v_{prpr}$ vanishes). Such $p=q$ terms do not exist in many-body perturbation theory at zero temperature
since the occupation number $n_p$ of a spin orbital in the unperturbed reference state of the zero-temperature formalism is either $0$ or $1$, 
so that $n_p(1-n_p) = 0$. It is, of course, not surprising that $\bar{n}_p(1-\bar{n}_p)$ does not vanish at nonzero temperature. But the point Kohn 
and Luttinger made is that $\bar{n}_p(1-\bar{n}_p)$ generally does not vanish as $T \rightarrow 0$; in fact \cite{KoLu60},
\begin{equation}
  \label{eq:17a} 
  \lim_{T \rightarrow 0} \beta \bar{n}_p(1-\bar{n}_p) = \delta(\varepsilon_p - \mu). 
\end{equation}
Hence, if there are one-particle states with orbital energies precisely equal to the chemical potential $\mu$, then those states 
give contributions to the energy $E$ that are not contained in the zero-temperature theory. 

However, in practice this can only happen if $\mu$ in the zero-temperature limit ends up in a continuum of orbital energies. This is the case when the 
noninteracting reference system has the orbital energy spectrum of a metal, i.e., when there is no energy gap between the highest occupied orbital and 
the lowest unoccupied orbital. For such a system, second-order perturbation theory generally diverges. (As mentioned earlier, nondegenerate 
Rayleigh-Schr\"{o}dinger perturbation theory is useful only if the unperturbed reference state is energetically isolated from other eigenstates of $\hat{H}_0$.) 
Therefore, if one focuses 
on potential applications where second-order perturbation theory does not diverge, then there is no Kohn--Luttinger conundrum. Specifically, zero-temperature 
many-body perturbation theory to finite order, say MP2, is most meaningful for closed-shell systems (such as band insulators in solid-state electronic-structure 
theory). In this case, the ground-state Hartree-Fock orbital-energy spectrum has a gap between the HOMO and the LUMO. By definition of the HOMO and the LUMO
for a closed-shell system, the chemical potential at $T=0$ lies inside this gap (see also Refs.~\cite{GyHa68,PePa82,Kapl06}). 
Hence, all terms depending on $\bar{n}_p(1-\bar{n}_p)$ vanish as $T \rightarrow 0$. For such systems, we may expect that in the limit $T \rightarrow 0$, 
FT-MBPT(2) connects smoothly to ZT-MBPT(2).

Using the fact that
\begin{equation}                                                                                                               
  \label{eq:20a} 
  \bar{n}_q(1-\bar{n}_p)\left\{1-e^{-\beta(\varepsilon_p - \varepsilon_q)}\right\} = \bar{n}_q - \bar{n}_p,
\end{equation}
it may be shown that the sum over $p$ and $q$ involving the factor $\left\{1-e^{-\beta(\varepsilon_p - \varepsilon_q)}\right\}/\left\{\varepsilon_p - \varepsilon_q\right\}$ in the first second-order term in Eq.~(\ref{eq:13}) vanishes (because the corresponding summand is antisymmetric in the indices $p$ and $q$). By a similar argument, the sum over $p$, $q$, $r$, and $s$ involving the factor $\left\{1-e^{-\beta(\varepsilon_p + \varepsilon_q - \varepsilon_r - \varepsilon_s)}\right\}/\left\{\varepsilon_p + \varepsilon_q - \varepsilon_r - \varepsilon_s\right\}$ in the second second-order term in Eq.~(\ref{eq:13}) vanishes as well.
Hence, through second order,
\begin{eqnarray}
  \label{eq:13a}
  \ln{Z_G} & = & \sum_p \ln{\left(1 + e^{-\beta(\varepsilon_p - \mu)}\right)} \\
  & & -\beta \left(\sum_p v_{pp} \bar{n}_p + \frac{1}{2} \sum_{p,q} v_{pq[pq]} \bar{n}_p \bar{n}_q\right) \nonumber \\
  & &  + \beta \sum_{p,q} \frac{\bar{n}_q(1-\bar{n}_p)}{\varepsilon_p - \varepsilon_q}
  \left|v_{pq} + \sum_{r} v_{pr[qr]} \bar{n}_r\right|^2 \nonumber \\          
  & & + \frac{1}{4} \beta \sum_{p,q,r,s} \frac{|v_{pq[rs]}|^2 \bar{n}_r \bar{n}_s 
    (1-\bar{n}_p) (1-\bar{n}_q)}{\varepsilon_p + \varepsilon_q - \varepsilon_r - \varepsilon_s}. \nonumber
\end{eqnarray}
Therefore, it follows from Eq.~(\ref{eq:16}) that, through second order, the grand potential is given by
\begin{eqnarray}
  \label{eq:13b}                                                                                                                
  \Omega & = & -\frac{1}{\beta}\sum_p \ln{\left(1 + e^{-\beta(\varepsilon_p - \mu)}\right)} \\
  & & + \sum_p v_{pp} \bar{n}_p + \frac{1}{2} \sum_{p,q} v_{pq[pq]} \bar{n}_p \bar{n}_q \nonumber \\         
  & & - \sum_{p,q} \frac{\bar{n}_q(1-\bar{n}_p)}{\varepsilon_p - \varepsilon_q}
  \left|v_{pq} + \sum_{r} v_{pr[qr]} \bar{n}_r\right|^2 \nonumber \\                                                            
  & & - \frac{1}{4} \sum_{p,q,r,s} \frac{|v_{pq[rs]}|^2 \bar{n}_r \bar{n}_s                                               
    (1-\bar{n}_p) (1-\bar{n}_q)}{\varepsilon_p + \varepsilon_q - \varepsilon_r - \varepsilon_s}. \nonumber                      
\end{eqnarray}
Note that the first- and second-order corrections to the grand potential are identical to the terms Hirata and He
identified, incorrectly, as the first- and second-order corrections to the energy (within what they called 
``conventional MBPT at a nonzero temperature''; cf. Sec.~II of Ref.~\cite{HiHe13}). In view of Eq.~(\ref{eq:15}),
it is obvious that $\Omega$ in Eq. (\ref{eq:13b}) does not converge to the ZT-MBPT(2) energy $E_0$ [cf. Eqs. (\ref{eq:2d})-(\ref{eq:2g})] 
as $T \rightarrow 0$.

\section{\label{sec:5} FT-MBPT(2) expressions for the mean energy and particle number}

The Kohn--Luttinger approach is suitable for investigating the zero-temperature limit. However, in order to obtain expressions 
for the mean particle number and, particularly, the mean energy of the 
interacting system at finite temperature, it is better to pursue the following strategy. Differentiating $\ln{Z_G}$ with respect to $\beta$ and $\mu$, respectively, gives
\begin{equation}
  \label{eq:18}
  \frac{\partial}{\partial \beta} \ln{Z_G} = -E + \mu \langle \hat{N} \rangle
\end{equation}
and 
\begin{equation}
  \label{eq:19}
  \frac{\partial}{\partial \mu} \ln{Z_G} = \beta \langle \hat{N} \rangle.
\end{equation}
When evaluating these expressions using Eq.~(\ref{eq:13a}), we assume that the eigenvalues and eigenstates of the unperturbed Hamiltonian, $\hat{H}_0$, are 
independent of the temperature and the chemical potential, so that the $\varepsilon_p$, $v_{pq}$, and $v_{pq[rs]}$ in Eq.~(\ref{eq:13a}) depend on neither 
$\beta$ nor $\mu$. (Otherwise, the situation would become considerably more complicated!)

In this way, we obtain for the mean particle number:
\begin{equation}
  \label{eq:21}
  \langle \hat{N} \rangle = \langle \hat{N} \rangle^{(0)} + \langle \hat{N} \rangle^{(1)} + \langle \hat{N} \rangle^{(2)} + \ldots,
\end{equation}
where
\begin{eqnarray}
  \label{eq:22}
  \langle \hat{N} \rangle^{(0)} & = & \sum_p \bar{n}_p, \\
  \label{eq:23}
  \langle \hat{N} \rangle^{(1)} & = & -\sum_p \bar{v}_{pp} \frac{\partial \bar{n}_p}{\partial \mu}, \\
  \label{eq:24}
  \langle \hat{N} \rangle^{(2)} & = & \sum_{p,q} \frac{1}{\varepsilon_p - \varepsilon_q} 
  \frac{\partial}{\partial \mu}\left[\bar{n}_q(1-\bar{n}_p)\left|\bar{v}_{pq}\right|^2\right] \\
  & & + \frac{1}{4} \sum_{p,q,r,s} \frac{|v_{pq[rs]}|^2 }{\varepsilon_p + \varepsilon_q - \varepsilon_r - \varepsilon_s}
  \frac{\partial}{\partial \mu}\left[\bar{n}_r \bar{n}_s (1-\bar{n}_p) (1-\bar{n}_q)\right], \nonumber
\end{eqnarray}
and
\begin{equation}
  \label{eq:24a}
  \bar{v}_{pq} = v_{pq} + \sum_{r} v_{pr[qr]} \bar{n}_r
\end{equation}
is an effective one-body-perturbation matrix element. 
The notation employed indicates for each $\langle \hat{N} \rangle^{(n)}$ the respective expansion order $n$ in the interaction, 
treating the chemical potential as an independent parameter (consistent with the spirit of the grand-canonical ensemble). Normally, 
however, $\langle \hat{N} \rangle$ (rather than $\mu$) is given. Then $\mu$ must be chosen such that Eq.~(\ref{eq:21}) 
is satisfied. In other words, Eq.~(\ref{eq:21}) must be solved for $\mu$ for every $\beta$, thus rendering the chemical potential a temperature- 
and $\langle \hat{N} \rangle$-dependent quantity. (Thus, one could, in principle, expand $\mu$ in orders of the interaction, but this is not done here.)

Note the appearance, in Eqs.~(\ref{eq:23}) and (\ref{eq:24}), of the partial derivative of Fermi-Dirac factors with respect to the chemical potential,
\begin{equation}
  \label{eq:25}
  \frac{\partial \bar{n}_p}{\partial \mu} = \beta \bar{n}_p(1-\bar{n}_p).
\end{equation}
This means that, in view of Eq.~(\ref{eq:17a}), $\langle \hat{N} \rangle^{(1)}$ and $\langle \hat{N} \rangle^{(2)}$ vanish in the zero-temperature 
limit---for those systems that are suitable for a finite-order perturbative treatment such as MP2.

When using Eq.~(\ref{eq:18}) to calculate the mean energy, one should note that 
\begin{equation}
  \label{eq:26}
  \frac{\partial \bar{n}_p}{\partial \beta} = \frac{\mu - \varepsilon_p}{\beta}\frac{\partial \bar{n}_p}{\partial \mu}.
\end{equation}
It is then easy to see that, when evaluating $\partial \ln{Z_G} / \partial \beta$, the factors $(\mu/\beta)\partial \bar{n}_p / \partial \mu$ contribute to 
$\mu \langle \hat{N} \rangle$; the factors $-(\varepsilon_p/\beta)\partial \bar{n}_p / \partial \mu$ contribute additional ``anomalous'' terms to the mean energy.
Overall, we find:
\begin{equation}
  \label{eq:27}
  E = E^{(0)} + E^{(1)}_{\Omega} + E^{(1)}_{\mathrm{A}} + E^{(2)}_{\Omega} + E^{(2)}_{\mathrm{A}} + \ldots,
\end{equation}
where
\begin{eqnarray}
  \label{eq:28}
  E^{(0)} & = & \sum_p \varepsilon_p \bar{n}_p, \\
  \label{eq:29}
  E^{(1)}_{\Omega} & = & \sum_p v_{pp} \bar{n}_p + \frac{1}{2} \sum_{p,q} v_{pq[pq]} \bar{n}_p \bar{n}_q, \\
  \label{eq:30}
  E^{(1)}_{\mathrm{A}} & = & -\sum_p \varepsilon_p \bar{v}_{pp} \frac{\partial \bar{n}_p}{\partial \mu}, \\
  \label{eq:31}
  E^{(2)}_{\Omega} & = & -\sum_{p,q} \frac{\bar{n}_q(1-\bar{n}_p)}{\varepsilon_p - \varepsilon_q}
  \left|\bar{v}_{pq}\right|^2 - \frac{1}{4} \sum_{p,q,r,s} 
  \frac{|v_{pq[rs]}|^2 \bar{n}_r \bar{n}_s (1-\bar{n}_p) (1-\bar{n}_q)}{\varepsilon_p + \varepsilon_q - \varepsilon_r - \varepsilon_s}, \\
  \label{eq:32}
  E^{(2)}_{\mathrm{A}} & = & \sum_{p,q} \frac{\left|\bar{v}_{pq}\right|^2}{\varepsilon_p - \varepsilon_q} 
  \left\{\varepsilon_q(1-\bar{n}_p)\frac{\partial \bar{n}_q}{\partial \mu} - \varepsilon_p \bar{n}_q \frac{\partial \bar{n}_p}{\partial \mu}\right\} \\
  & & + 2 \sum_{p,q} \frac{\bar{n}_q (1-\bar{n}_p)}{\varepsilon_p - \varepsilon_q}
  \operatorname{Re}\left\{\left(\sum_{r} v_{pr[qr]}^{\ast}\varepsilon_r \frac{\partial \bar{n}_r}{\partial \mu}\right) \bar{v}_{pq}\right\} \nonumber \\
  & & + \frac{1}{4} \sum_{p,q,r,s} \frac{|v_{pq[rs]}|^2 (1-\bar{n}_p) (1-\bar{n}_q)}{\varepsilon_p + \varepsilon_q - \varepsilon_r - \varepsilon_s} 
  \left\{\varepsilon_r \frac{\partial \bar{n}_r}{\partial \mu}\bar{n}_s + \varepsilon_s \bar{n}_r \frac{\partial \bar{n}_s}{\partial \mu}\right\} \nonumber \\ 
  & & - \frac{1}{4} \sum_{p,q,r,s} \frac{|v_{pq[rs]}|^2 \bar{n}_r \bar{n}_s}{\varepsilon_p + \varepsilon_q - \varepsilon_r - \varepsilon_s}
  \left\{\varepsilon_p \frac{\partial \bar{n}_p}{\partial \mu} (1-\bar{n}_q)
    + \varepsilon_q (1-\bar{n}_p) \frac{\partial \bar{n}_q}{\partial \mu}\right\}. \nonumber
\end{eqnarray}

The zeroth-order contribution, Eq.~(\ref{eq:28}), is simply the sum over all orbital energies weighted by the corresponding orbital occupation numbers.
We have separated the first- and second-order corrections into terms that are already present in the perturbation expansion of the grand 
potential---Eqs.~(\ref{eq:29}) and (\ref{eq:31}) [cf. Eq.~(\ref{eq:13b})]---and those that are new---Eqs.~(\ref{eq:30}) and (\ref{eq:32}). 
(The latter two are missing in what Hirata and He incorrectly call ``conventional MBPT at a nonzero temperature'' \cite{HiHe13}.)
At nonzero temperature, all these corrections have to be taken into consideration, within second-order many-body perturbation theory.
As the temperature goes to zero, under the conditions discussed in Sec.~\ref{sec:4} $E^{(0)}$, $E^{(1)}_{\Omega}$, and $E^{(2)}_{\Omega}$
converge to the zeroth-, first-, and second-order contributions, respectively, expected from ZT-MBPT(2), 
and $E^{(1)}_{\mathrm{A}}$ and $E^{(2)}_{\mathrm{A}}$ vanish.

For some applications, it may be of interest to combine Eqs.~(\ref{eq:15}), (\ref{eq:13b}), (\ref{eq:21})-(\ref{eq:24a}), and 
(\ref{eq:27})-(\ref{eq:32}) to construct the FT-MBPT(2) approximation to the electronic entropy, $S$.

\section{\label{sec:6} Remarks on ``renormalized MBPT at a nonzero temperature''}

In Ref.~\cite{HiHe13}, Hirata and He take the view that the Kohn--Luttinger conundrum 
indicates the need for a specific modification of the conventional finite-temperature perturbation theory, referred to as ``renormalized MBPT at a nonzero temperature.''

To that end, they adopt the use of normal-ordered operator products as introduced by Matsubara for quantum many-body problems at finite temperature 
\cite{Mats55,Thou57}. These 
products are defined such that the corresponding expectation value for the noninteracting statistical ensemble vanishes. For example, let the normal-ordered product of 
$\hat{c}^{\dag}_p \hat{c}_q$ be
\begin{equation}
\label{eq:33}
  \{\hat{c}^{\dag}_p \hat{c}_q\} = \hat{c}^{\dag}_p \hat{c}_q - \delta_{pq} \bar{n}_p.
\end{equation}
Then, 
\begin{equation}
\label{eq:34}
  \langle \{\hat{c}^{\dag}_p \hat{c}_q\} \rangle_0 = \mathrm{Tr}\left\{\hat{\rho}_0 \{\hat{c}^{\dag}_p \hat{c}_q\} \right\} = 0.
\end{equation}
Using the normal-ordered operator products, 
the operators $\hat{H}_0$ and $\hat{H}_1$ [Eqs.~(\ref{eq:2a}) and (\ref{eq:2b})]
can be re-written accordingly.
For example, Eq.~(81) in Ref.~\cite{HiHe13} gives the unperturbed Hamiltonian in normal order. Supposing that 
$\hat{H}_0$ is diagonal, its normal-ordered form reads, in our notation,
\begin{equation}
\label{eq:35}
\hat{H}_0 = \sum_p \varepsilon_p \bar{n}_p + \sum_p \varepsilon_p \{\hat{c}^{\dag}_p \hat{c}_p\}.
\end{equation}
Of course, using Eq.~(\ref{eq:33}), the original form~(\ref{eq:2a}) of $\hat{H}_0$
can readily be restored. 
So it is important to note that normal-ordering does not change operators such as the parts 
$\hat{H}_0$ and $\hat{H}_1$ of the Hamiltonian---it merely brings them into a form that is advantageous in the evaluation of thermal averages. In particular,
even though the normal-ordered forms may suggest otherwise, there results no temperature dependence whatsoever.  

The temperature dependence comes into play via the entity $\Phi_0$, referred to, somewhat vaguely and only in the Appendix of Ref.~\cite{HiHe13}, 
as ``the HF reference wave function at nonzero temperature.'' Unfortunately, there is no further characterization of $\Phi_0$.
However, it may be inferred from the zeroth-order energy (see Eq.~(79)
in Ref.~\cite{HiHe13}) obtained according to 
\begin{equation}
E_{\mathrm{R}}^{(0)} = \langle \Phi_0 |\hat{H}_0| \Phi_0 \rangle = 
\sum_p \varepsilon_p \bar{n}_p
\end{equation}
that an expression of the form $\langle \Phi_0 |\dots| \Phi_0 \rangle$ appears to 
be a symbolic notation for $\mathrm{Tr}\left\{\hat{\rho}_0 ... \right\}$. 

While using an alternative formalism is of course perfectly legitimate, the actual 
deviation from the conventional approach is a problematic modification
of the second-order corrections to the energy. This modification is expressed in
Eqs.~(84) and (85) of Ref.~\cite{HiHe13}, reading, in original form,
\begin{eqnarray}
  \label{eq:36}
  E_{\mathrm{R}}^{(2N)} & = & \frac{1}{4} \sum_{p,q,r,s} \frac{\langle \Phi_0 |\hat{H}_1| \Phi_{rs}^{pq} \rangle 
    \langle \Phi_{rs}^{pq} |\hat{H}_1| \Phi_0 \rangle}{\langle \Phi_0 |\hat{H}_0| \Phi_0 \rangle - \langle \Phi_{rs}^{pq} |\hat{H}_0| \Phi_{rs}^{pq} \rangle}, \\
  \label{eq:37}
  E_{\mathrm{R}}^{(2A)} & = & \sum_{p,q} \frac{\langle \Phi_0 |\hat{H}_1| \Phi_{q}^{p} \rangle
    \langle \Phi_{q}^{p} |\hat{H}_1| \Phi_0 \rangle}{\langle \Phi_0 |\hat{H}_0| \Phi_0 \rangle - \langle \Phi_{q}^{p} |\hat{H}_0| \Phi_{q}^{p} \rangle}.
\end{eqnarray}
In Ref.~\cite{HiHe13}, $\Phi_{q}^{p}$ and  $\Phi_{rs}^{pq}$ are referred to as ``one- and two-electron excited eigenstates of $\hat{H}_0$.''
However, this is not quite accurate as the actual usage of these 'states' derives from applying appropriate operator products to $\Phi_0$,
e.g.,  $|\Phi_{q}^{p} \rangle = \hat{c}^{\dag}_p\hat{c}_q |\Phi_0 \rangle$.

In striking contrast to conventional FT-MBPT(2) expressions, the latter 'renormalized'
equations imply thermal averaging not only in the numerators but, independently, in the denominators as well. The final results are given by Eqs.~(89) and (93) 
in Ref.~\cite{HiHe13}; using our notation the latter equations take on the form 
\begin{align}
\label{eq:hho2a}
E_{\mathrm{R}}^{(2N)} =& - \frac{1}{4} \sum_{p,q,r,s} 
\frac{v_{pq[rs]} v_{rs[pq]}\bar{n}_r\bar{n}_s (1-\bar{n}_p)(1-\bar{n}_q)}
{\varepsilon_p (1-\bar{n}_p) +\varepsilon_q (1-\bar{n}_q) -\varepsilon_r \bar{n}_r - \varepsilon_s \bar{n}_s}, \\
\label{eq:hho2b}
E_{\mathrm{R}}^{(2A)} =& -\sum_{p,q} \frac{\bar{n}_q(1-\bar{n}_p)}
{\varepsilon_p (1-\bar{n}_p) - \varepsilon_q\bar{n}_q} \left|\bar{v}_{pq}\right|^2.
\end{align}
Note that in the second line we have used the more general matrix element $\bar{v}_{pq}$ as given by Eq.~(\ref{eq:24a}),
rather than $\langle p||q \rangle$ defined by Eqs.~(18)-(20) in Ref.~\cite{HiHe13}.

In the latter form, Eqs.~(\ref{eq:hho2a}) and (\ref{eq:hho2b}) can directly be compared to
the rigorous FT-MBPT(2) results given by Eqs.~(\ref{eq:31}) and (\ref{eq:32}). 
Unsurprisingly, the contributions $E^{(2)}_{\mathrm{A}}$ of Eq.~(\ref{eq:32}), 
arising from derivatives with respect to $\mu$, are completely absent in the renormalized MBPT version. 
More interesting is the comparison of the two terms on the right-hand side of Eq.~(\ref{eq:31})
with their respective counterparts in Eqs.~(\ref{eq:hho2a}) and (\ref{eq:hho2b}). 
The only if significant difference is the presence of Fermi-Dirac factors in the 
denominators of the renormalized FT-MBPT(2) expressions. 
This deviation from the exact result shows that the renormalized FT-MBPT(2), as reflected by  
 Eqs.~(\ref{eq:36}, \ref{eq:37}), must be seen as a mere \textit{ad hoc}
modification, lacking a rigorous foundation.  

Of course, this does not rule out the possibility that the renormalized 
variant of FT-MBPT may prove useful or even superior to conventional FT-MBPT 
in specific applications, as a recent study of one-dimensional solids by He 
et al. \cite{HeRy14} seems to indicate. However, in view of the analysis given 
here, a systematic improvement eventually afforded by the renormalized FT-MBPT 
scheme cannot simply be taken for granted but should rather be seen in need of 
explanation.

\section{\label{sec:7} Conclusions}

We have presented an algebraic, nondiagrammatic derivation of FT-MBPT(2) using a theoretical framework that, we believe, is accessible to chemical
physicists familiar with the formalism of second quantization (at the level of standard textbooks such as Ref.~\cite{SzOs96}). Our focus has been on
{\em fermionic} many-body systems (such as electrons), but apart from that, our results are general, i.e., the equations presented may be employed for,
in principle, arbitrary one- and two-body interactions.

In practice, however, the usefulness of FT-MBPT(2) depends on whether corrections beyond second order are important or not. This is analogous to asking
whether, say, MP2 is sufficient for describing an interacting many-electron system. Since ZT-MBPT(2) is simply nondegenerate second-order
Rayleigh-Schr\"{o}dinger perturbation theory for a Hamiltonian containing one- and two-body operators (the unperturbed part being a one-body operator), 
it is clear that ZT-MBPT(2) will be accurate only if the reference state selected is energetically isolated from all other eigenstates of the unperturbed 
part of the Hamiltonian. This condition is only met for systems with a nonzero (ideally, large) HOMO--LUMO gap.

The equations for the mean particle number, Eqs.~(\ref{eq:21})-(\ref{eq:24a}), and the equations for the mean energy, Eqs.~(\ref{eq:27})-(\ref{eq:32}),
may be considered the central results of FT-MBPT(2). In the presence of interactions not contained in the unperturbed part of the Hamiltonian, 
both sets of equations contain terms that depend on factors of the type $\partial \bar{n}_p / \partial \mu$, which is proportional to 
$\bar{n}_p(1-\bar{n}_p)$. These terms are not expected from the perspective of ZT-MBPT(2) (in contrast to what has been suggested in the literature,
FT-MBPT is not simply ZT-MBPT with fractional occupation numbers \cite{KoTa15}). In fact, as Kohn and Luttinger pointed
out \cite{KoLu60}, these terms do not, in general, vanish when taking the limit $T \rightarrow 0$---and, hence, indicate a formal disagreement with ZT-MBPT(2).
But this is a problem only for systems for which second-order perturbation theory may not be expected to provide reliable results anyway.

Nevertheless, one may ask: Why is it that FT-MBPT(2) does not generally converge to standard MBPT(2) in the zero-temperature limit? Let us ask a different,
but closely related question: Should we expect the density operator $\hat{\rho}$ [Eq.~(\ref{eq:1})] of the grand-canonical ensemble to go over, in the
zero-temperature limit, into the pure-state density operator $|\Psi_0^N\rangle \langle\Psi_0^N|$, where $|\Psi_0^N\rangle$ is the ground state of 
$\hat{H}$ (assumed, for simplicity, to be nondegenerate) for a {\em specific} particle number, $N$? In other words, do particle-number fluctuations
vanish, in the grand-canonical ensemble, as $T \rightarrow 0$? 

The answer is no: The particle number is generally not conserved. This is easy to see by calculating the square of the uncertainty of the particle number
within the noninteracting grand-canonical ensemble:
\begin{eqnarray}
  \label{eq:40}
  (\Delta N)^2 & = & \langle \hat{N}^2 \rangle_0 - \langle \hat{N} \rangle_0^2 \\
  & = & \sum_{p,q} \langle \hat{n}_p \hat{n}_q \rangle_0 - \sum_{p,q} \langle \hat{n}_p \rangle_0 \langle \hat{n}_q \rangle_0 \nonumber \\
  & = & \sum_p \left\{\bar{n}_p - \bar{n}_p^2\right\} + \sum_{p \ne q} \left\{\bar{n}_p \bar{n}_q - \bar{n}_p \bar{n}_q\right\} \nonumber \\
  & = & \sum_p \bar{n}_p(1-\bar{n}_p). \nonumber
\end{eqnarray}
We note the appearance of $\bar{n}_p(1-\bar{n}_p)$, which remains nonzero, even as $T \rightarrow 0$, for all orbitals whose energy $\varepsilon_p$ equals
the chemical potential, $\mu$ ($\bar{n}_p = 1/2$ for $\varepsilon_p = \mu$). Thus, if there is no HOMO--LUMO gap in the noninteracting reference system, 
then the zero-temperature limit of 
the grand-canonical ensemble does not lead to a state with perfectly well-defined particle number. This
is the ultimate reason why zero-temperature MBPT(2) and FT-MBPT(2) at zero temperature differ for such systems.

\section{Acknowledgments}
R.S. would like to thank Beata Ziaja, Sang-Kil Son, and Ludger Inhester for inspiring discussions.

\end{document}